\documentclass[
aps,prl,amsmath,amssymb,twocolumn,
floatfix,groupedaddress]{revtex4-2}
\usepackage{mathrsfs}
\usepackage{color}
\usepackage{graphicx}
\usepackage{dcolumn}
\usepackage{braket}
\usepackage{bm}
\usepackage{hyperref}
\usepackage{hypernat}
\usepackage{version}
\usepackage{epstopdf}
\usepackage{float}

\usepackage[normalem]{ulem} 
\DeclareGraphicsExtensions{.eps,.ps,.png}

\DeclareMathOperator*{\SumInt}{%
\mathchoice%
  {\ooalign{$\displaystyle\sum$\cr\hidewidth$\displaystyle\int$\hidewidth\cr}}
  {\ooalign{\raisebox{.14\height}{\scalebox{.7}{$\textstyle\sum$}}\cr\hidewidth$\textstyle\int$\hidewidth\cr}}
  {\ooalign{\raisebox{.2\height}{\scalebox{.6}{$\scriptstyle\sum$}}\cr$\scriptstyle\int$\cr}}
  {\ooalign{\raisebox{.2\height}{\scalebox{.6}{$\scriptstyle\sum$}}\cr$\scriptstyle\int$\cr}}
}

\begin{document}

\preprint{APS/123-QED}
\title{Time-dependent Hole States in Multiconfigurational Time-Dependent Hartree-Fock Approaches: A Time-Domain Generalization of Extended Koopmans' Theorem}
\author{Zhao-Han Zhang$^{1,2}$}
\author{Yang Li$^{1}$}\email{liyang22@sjtu.edu.cn}
\author{Himadri Pathak$^{4,5}$}
\author{Takeshi Sato$^{6,7,8}$}
\author{Kenichi L. Ishikawa$^{6,7,8,9}$}
\author{Feng He$^{1,3}$}\email{fhe@sjtu.edu.cn}

\affiliation{
$^1$Key Laboratory for Laser Plasmas (Ministry of Education) and School of Physics and Astronomy,
Collaborative Innovation Center for IFSA (CICIFSA), Shanghai Jiao Tong University, Shanghai 200240, China\\
$^2$Max-Planck-Institut f\"{u}r Kernphysik, Saupfercheckweg 1, 69117 Heidelberg, Germany\\
$^3$Tsung-Dao Lee Institute, Shanghai Jiao Tong University, Shanghai 201210, China\\
$^4$Quantum Mathematical Science Team, Division of Applied Mathematical Science, RIKEN Center for Interdisciplinary Theoretical and Mathematical Sciences (iTHEMS),  2-1 Hirosawa Wako, Saitama 351-0198, Japan\\
$^5$Computational Molecular Science Research Team, RIKEN Center for Computational Science (R-CCS), 7-1-26 Minatojima-minami-machi, Chuo-ku, Kobe, Hyogo 650-0047, Japan\\
$^6$Department of Nuclear Engineering and Management, Graduate School of Engineering, The University of Tokyo,7-3-1 Hongo, Bunkyo-ku, Tokyo 113-8656, Japan\\
$^7$Photon Science Center, Graduate School of Engineering, The University of Tokyo, 7-3-1 Hongo, Bunkyo-ku, Tokyo 113-8656, Japan\\
$^8$Research Institute for Photon Science and Laser Technology, The University of Tokyo, 7-3-1 Hongo, Bunkyo-ku, Tokyo 113-0033, Japan\\
$^9$Institute for Attosecond Laser Facility, The University of Tokyo, 7-3-1 Hongo, Bunkyo-ku, Tokyo 113-0033, Japan
}
\date{\today}

\begin{abstract}
We introduce a framework for resolving electron-hole dynamics within wavefunction-based multiconfigurational time-dependent Hartree-Fock (MCTDHF) theory. Central to this framework is a time-domain generalization of the extended Koopmans' theorem, which rigorously defines time-dependent hole states through single-electron removal. From this foundation, we prove the existence of exact equations of motion for time-dependent Dyson orbitals, enabling instantaneous construction of photofragments' reduced density matrices. The formalism further yields a systematic procedure to extract hole-resolved observables, such as channel-resolved photoelectron momentum distributions, directly from time-dependent \textit{ab initio} wavefunctions. As a demonstration, we employ an attosecond $\omega-2\omega$ laser strategy to control hole dynamics, thereby resolving a long-standing challenge in MCTDHF simulations. This advance opens a pathway for exploring correlated multielectron dynamics in atoms and molecules under ultrafast laser fields.
\end{abstract}

%\pacs{42.50.Hz 42.65.Re 82.30.Lp}

\maketitle
Koopmans’ theorem, a cornerstone of electronic structure theory, states that within Hartree-Fock, the ionization energies equal the negative of the occupied orbital energies~\cite{koopmans1934zuordnung}. To incorporate electron correlation, Hartree-Fock has been extended to multiconfigurational frameworks, giving rise to the \textit{extended Koopmans’ theorem} (EKT)~\cite{day1974generalization,smith1975extension,morrell1975calculation,vanfleteren2009exact}. It provides more accurate ionization energies and electron affinities and naturally yields \textit{Dyson orbitals} (hole wavefunctions)~\cite{smith1975extension}, which encode amplitude- and phase-resolved information about ionization channels beyond Hartree-Fock, enabling a direct quantitative link between correlated electronic structure methods and spectroscopic observables.

Inheriting from Hartree-Fock theory, the multiconfigurational time-dependent Hartree-Fock (MCTDHF) method~\cite{lode2020colloquium,hochstuhl2014time,ishikawa2015review,zanghellini2003mctdhf,kato2004time,caillat2005correlated} and its variants~\cite{sato2013time,miyagi2013time,sato2015time} have become powerful tools for solving the multielectron time-dependent Schr\"{o}dinger equation (ME-TDSE) beyond the perturbative regime. Over the past decade, MCTDHF has enabled quantitatively accurate simulations of ultrafast ionization and high-harmonic generation across various laser conditions and multielectron targets~\cite{nest2007time,remacle2007laser,kato2008time,nest2008pump,hochstuhl2011two,haxton2012single,hochstuhl2014time,ohmura2014characterization,haxton2014ultrafast,li2016direct,sato2016time,sawada2016implementation,liao2017probing,tikhomirov2017high,greenman2017optimized,omiste2018attosecond,orimo2018implementation,li2019high,orimo2019application,lotstedt2020excited,lotstedt2020static,li2021implementation,omiste2021photoionization,orimo2023use}. Yet, despite the foundational role of EKT in stationary electronic structure theory, its rigorous counterpart in time-dependent theories remains absent.

The motivation for such a generalization arises from the complexity of photofragmentation processes, such as photoionization and photodissociation, triggered by attosecond lasers from high-harmonic sources~\cite{sansone2006isolated,chini2014generation,gaumnitz2017streaking,wang2024ultrashort} and X-ray free electron lasers~\cite{huang2017generating,duris2020tunable}.
In these scenarios, the fragments are compound systems with intricate internal structures that are difficult for detection, since most internal states (rotational, vibrational, and electronic) soon decay before reaching the detector, leaving primarily kinetic observables (e.g., asymptotic momenta of photoelectrons and photoions). 
The broad bandwidth of attosecond pulses further complicates the interpretation, as numerous ionization channels overlap in the spectra~\cite{li2024attosecond,rodriguez2024core,ishikawa2023control}. Consequently, hole dynamics often follows attosecond ionization~\cite{breidbach2005,PhysRevLett.132.263202,science.1254061,science.aab2160}. Such overlap may also induce intriguing phenomena profoundly associated with quantum entanglement and coherence~\cite{vrakking2021control,koll2022experimental,laurell2025measuring}. Even with advanced techniques such as COLTRIMS~\cite{dorner2000cold}, disentangling contributions from different internal states remains challenging, particularly when the ion evolves in time, such as laser-driven hole dynamics in strong laser fields~\cite{calegari2014ultrafast,kraus2015measurement,he2022filming}. The need for generalization is further underscored in strong-field ionization, where electrons from multiple orbitals may be released~\cite{kanai2005quantum,mcfarland2008high} or where channel coupling effects emerge~\cite{shu2022channel}. These considerations necessitate a theory that retains the internal information of the fragments while tracking hole dynamics in real time. Unfortunately, this remains a long-standing challenge for MCTDHF, which ultimately stems from the absence of well-defined time-dependent hole states (TDHSs).

In this Letter, we establish a time-domain generalization of EKT and introduce a rigorous formulation of the TDHS within the MCTDHF framework. We first specify three essential \textit{conditions} that a physically \textit{meaningful} TDHS must satisfy. Constrained by these \textit{conditions}, we propose an \textit{ansatz} of the TDHS and derive its equation of motion (EOM). We demonstrate that the stationary limit of this equation recovers the conventional EKT, validating its time-domain generalization. We further prove the existence of an \textit{exact} EOM of the time-dependent Dyson orbital, a fundamental property of ME-TDSE beyond any approximated framework. This formalism provides a physically rigorous definition of channel-resolved observables, correcting earlier misinterpretations based on time-dependent natural orbitals~\cite{ohmura2020analysis,ohmura2022investigation}, and enabling channel-resolved extensions of widely used analysis techniques in ultrafast physics. By applying this capability, we uncover a new physical scenario in which attosecond $\omega$-$2\omega$ pulses coherently steer hole dynamics in dimer systems, demonstrating controllable charge migration on ultrafast timescales. This finding illustrates how the formalism not only resolves a conceptual challenge but also opens a route to explore correlated electron dynamics.

A central difficulty in defining TDHS is that, in MCTDHF, all neutral and ionic states are encoded within a single correlated \textit{ab initio} full $N$-electron wavefunction $(\Psi)$, expanded in terms of a single set of time-dependent orbitals $(\phi_i)$ and configuration expansion coefficients $(C_I)$,
\begin{equation}
\label{Def:Psi}%%%
|\Psi(t)\rangle=\sum_I C_I(t)|\phi_{I_1}(t)\phi_{I_2}(t)\cdots\phi_{I_N}(t)\rangle.
\end{equation}
From this representation, the full photoelectron momentum distribution (PMD) can be conveniently determined by the expectation value of the momentum density operator $(\hat{a}^\dag_{\bm{k}\sigma}\hat{a}_{\bm{k}\sigma})$ in the asymptotic region after a sufficiently long time ($t>t_f$),
\begin{equation}
\label{Def:PMD}%%%
\mathcal{P}(\bm{k}\sigma)=\langle\Psi(t_f)|\hat{a}^\dag_{\bm{k}\sigma}\hat{a}_{\bm{k}\sigma}|\Psi(t_f)\rangle.
\end{equation}
However, extracting the residual ion state correlated with the released photoelectron is conceptually challenging. Constructing a single-vacancy state from Eq.~\eqref{Def:Psi} requires removing a time-dependent orbital, which may deviate significantly from stationary solutions, particularly in the non-perturbative regime. Even worse, the ion states themselves evolve dynamically, further departing from the stationary forms. We now conquer this challenge.

As all presented quantities depend on time, we omit the time variable unless we refer to a specific instant. We define $\bm{\kappa}\equiv(\bm{k}\sigma)$ as a collective coordinate for momentum $\bm{k}$ and spin $\sigma$. The single-electron state $|\bm{\kappa}\rangle\equiv\hat{a}^\dag_{\bm{\kappa}}|vac\rangle$ represents an electron in the asymptotic region with kinetic momentum $\bm{k}$ and spin $\sigma$. Let $|\gamma\rangle$ be a general $(N-1)$-electron state that we are looking for, and $|\gamma,\bm{\kappa}\rangle\equiv\hat{a}^\dag_{\bm{\kappa}}|\gamma\rangle$ be the state with one extra electron in the asymptotic region. The channel-resolved projection $\mathcal{Q}_\gamma(\bm{\kappa})\equiv\langle\gamma,\bm{\kappa}|\Psi\rangle$ can be rigidly expressed as a linear combination of $\phi_\mu(\bm{\kappa})\equiv\langle\bm{\kappa}|\phi_\mu\rangle$, yielding (Einstein's summation convention is applied for Greek subscripts)
\begin{equation}\label{Def:Qgamma}%%%
\mathcal{Q}_\gamma(\bm{\kappa}) =  D_{\gamma\mu}\phi_\mu(\bm{\kappa}),
\end{equation}
where the coefficients $D_{\gamma\mu} \equiv \langle\gamma|\hat{a}_\mu|\Psi\rangle$ are independent of $\bm{\kappa}$ and nonzero only when $\phi_\mu$ is occupied. The quantity $\mathcal{P}_\gamma(\bm{\kappa})=|\mathcal{Q}_\gamma(\bm{\kappa})|^2$ is interpreted as channel-resolved PMD, accordingly.

Given these definitions, we introduce the first \textit{condition}: in the limit $t\rightarrow \infty$, the interaction between the photoelectron and the residual $(N-1)$-electron system vanishes, and hence $\mathcal{P}_\gamma(\bm{\kappa})$ should eventually be stable, formally expressed as
\begin{equation}
\lim_{t\rightarrow+\infty}\dot{\mathcal{P}}_\gamma(\bm{\kappa})=0,
\label{Condition:Conservation}
\end{equation}
which we refer to as the \textit{unitarity condition}. This condition applies not only to single ionization, but also to double or multiple ionization, where the residual system itself could be excited or ionized. An improper choice of $|\gamma\rangle$, such as a linear combination of two valid choices, could induce indefinite oscillations of $\mathcal{P}_{\gamma}$.

The second \textit{condition} stems from the requirement that the sum of $\mathcal{P}_\gamma(\bm{\kappa})$ over all possible $|\gamma\rangle$ must equal the total PMD, $\mathcal{P}(\bm{\kappa})=\phi^*_\mu(\bm{\kappa})\rho_{\mu\nu}\phi_\nu(\bm{\kappa})$, where the one-body reduced density matrix $\rho_{\mu\nu}=\langle\Psi|\hat{a}^\dag_\mu\hat{a}_\nu|\Psi\rangle$. Consequently, the valid choices of $|\gamma\rangle$ are limited by $\rho$. They should satisfy the following two equivalent equalities,
\begin{subequations}
\begin{align}
\label{Condition:GSRD}
\rho_{\mu\nu} &=  D^*_{\gamma\mu}D_{\gamma\nu}, \\
\label{Condition:Completeness}
\langle\Psi|\hat{a}^\dag_\mu\hat{a}_\nu|\Psi\rangle &= \langle\Psi|\hat{a}^\dag_\mu|\gamma\rangle\langle\gamma|\hat{a}_\nu|\Psi\rangle.
\end{align}
\end{subequations}
We refer to them as the \textit{sum condition}. Eq.~\eqref{Condition:GSRD} requires that $\rho=D^\dag D$. Hence, $D$ must be one of the generalized matrix square root decompositions of $\rho$. However, some decompositions are not \textit{meaningful}, as illustrated by an example later. Eq.~\eqref{Condition:Completeness} further implies that any single-vacancy state $\hat{a}_\mu|\Psi\rangle$ should lie entirely within the $(N-1)$-electron Hilbert space spanned by $|\gamma\rangle$, which we denote by $\varPi$.

The third \textit{condition} concerns the physical reality of $\mathcal{P}_\gamma$. In MCTDHF, a redundancy arises from the fact that two different sets of $\{\phi_i,C_I\}$, related by a unitary transformation, represent the same physical state. Typically, constraints on $R_{\mu\nu}\equiv i\langle\phi_\mu|\dot{\phi}_\nu\rangle$ eliminate this redundancy~\cite{meyer1990multi,caillat2005correlated}. A \textit{meaningful} $\mathcal{P}_\gamma$ must be independent of the choice of $R$, which we refer to as the \textit{physical condition}.

Guided by the \textit{conditions}, we expand the TDHS $|\gamma\rangle$ as (atomic units are used throughout)
\begin{equation}
\label{Def:TDHS}%%%
|\gamma\rangle = \hat{a}_\mu|\Psi\rangle Z_{\mu\gamma}.
\end{equation}
The states $\hat{a}_\mu|\Psi\rangle$ are directly available from MCTDHF, while the coefficient matrix $Z$ remains to be determined.
Eq.~\eqref{Def:TDHS} implies that $|\gamma\rangle$ satisfies the projected $(N-1)$-electron ME-TDSE:
\begin{equation}
\label{ProjectedTDSE}%%%
i\hat{\varPi}|\dot{\gamma}\rangle = \hat{\varPi}\hat{H}|\gamma\rangle,
\end{equation}
where $\hat{\varPi}$ is the projector of the subspace $\varPi$, and $\hat{H}$ is the time-dependent Hamiltonian that incorporates electron-electron, electron-nuclei and electron-laser interactions. Eq.~\eqref{ProjectedTDSE} ensures the \textit{physical condition}, since $\hat{\varPi}$ is invariant under unitary transformations of the orbitals. In the full configuration interaction limit, $\hat{\varPi}$ approaches the identity operator in the single-vacancy space and $|\gamma\rangle$ converges to the \textit{exact} $(N-1)$-electron ME-TDSE solutions. Combining Eqs.~\eqref{Def:TDHS} and~\eqref{ProjectedTDSE} yields an EOM for $Z$
\begin{subequations}
\begin{align}
\label{EOM:TDHS}%%%
-\tilde{F}^T Z&=i\rho \dot{Z},\\
\tilde{F}_{\mu\nu}&=(F_{\mu\tau}-R_{\mu\tau})\rho_{\nu\tau},
\end{align}
\end{subequations} with $F_{\mu\nu}\equiv\langle\phi_\mu|\hat{F}|\phi_\nu\rangle$, where $\hat{F}$ is the generalized Fock operator in MCTDHF~\cite{sato2013time}. Generally, the hole subspace evolves in time since $\hat{\varPi}|\dot{\gamma}\rangle\ne|\dot{\gamma}\rangle$ and $d\hat{\varPi}/dt\ne0$. 
 
The present definition exhibits several desirable properties required for a successful time-domain generalization of EKT. 
First, once a field-free stationary MCTDHF solution is reached, $\tilde{F}$ becomes real and symmetric, and the stationary condition of Eq.~\eqref{EOM:TDHS}, $i\dot{Z}=Z\mathcal{E}$, reduces to a generalized eigenvalue problem,
\begin{equation}
-\tilde{F}Z=\rho Z\mathcal{E},
\label{Koopmans}%%%
\end{equation}
which reproduces the static EKT. $|\gamma\rangle$ therefore coincides with the field-free hole states at $t=0$, and the overlap between $\gamma(0)$ and $\Psi(0)$ yields the static Dyson orbitals. The orthonormality, $\langle\gamma|\gamma'\rangle=(Z^\dag)_{\gamma\mu} \rho_{\mu\nu} Z_{\nu\gamma'}=\delta_{\gamma\gamma'}$, is satisfied at all times, as $Z^\dag \rho Z=I$ is initially satisfied and preserved during propagation.
Second, the \textit{sum condition} of Eq.~\eqref{Condition:GSRD} is fulfilled at any time, since
\begin{equation}
\label{Relation:D&Z}
D_{\gamma\mu}=\langle\gamma|\hat{a}_\mu|\Psi\rangle=(Z^\dag)_{\gamma\nu}\rho_{\nu\mu}=(Z^{-1})_{\gamma\mu},
\end{equation}
implying $D^\dag D=\rho^\dag Z Z^{-1}=\rho$.
This property may seem surprising because $\rho$ includes contributions from single ionization and multiple ionization, whereas the TDHS, composed of single-vacancy states, appears to describe single ionization only. In fact, a time-dependent ion can continue to ionize into higher charge states. As a result, $|\gamma\rangle$ includes not only singly charged ion states, but also all fragment states created from them. 
Third, and perhaps most importantly, the \textit{unitarity condition} is satisfied. To see this, we introduce a practical realization of $|\bm{\kappa}\rangle=\hat{\varTheta}|\chi_{\bm{\kappa}}\rangle$, where $\hat{\varTheta}$ is a mask operator to remove all wavepackets inside a radius $r_0$, and $\chi_{\bm{\kappa}}$ is a Volkov state satisfying $\hat{H}_V|\chi_{\bm{\kappa}}\rangle=i|\dot{\chi}_{\bm{\kappa}}\rangle$. Here, $\hat{H}_V$ is the Volkov Hamiltonian, and $\chi_{\bm{\kappa}}$ evolves into a plane wave with momentum $\bm{k}$ and spin $\sigma$ after the external field is terminated. The EOMs of $D_{\gamma\mu}$ and $\phi_\mu(\bm{\kappa})$ then follow,
\begin{subequations}
\begin{align}
\label{EOM:OrbMom}%%%
i\dot{\phi}_\mu(\bm{\kappa}) &= J_\mu(\bm{\kappa}) + \phi_\nu(\bm{\kappa})(R_{\nu\mu}-F_{\nu\mu}),\\
\label{EOM:D}%%%
i\dot{D}_{\gamma\mu} &= (F_{\mu\nu}-R_{\mu\nu})D_{\gamma\nu},
\end{align}
\end{subequations}
where $J_\mu(\bm{\kappa})$ is the time-dependent surface flux,
\begin{equation}
\begin{aligned}
J_\mu(\bm{\kappa})
&=\langle\chi_{\bm{\kappa}}|\hat{\varTheta}\hat{F}-\hat{H}_V\hat{\varTheta}|\phi_\mu\rangle \\
&\approx \langle\chi_{\bm{\kappa}}|[\hat{\varTheta},\hat{H}_V]|\phi_\mu\rangle.
\end{aligned}
\end{equation}
Using Eqs.~\eqref{EOM:OrbMom} and~\eqref{EOM:D}, the EOM of $\mathcal{Q}_{\gamma}(\bm{\kappa})$ is simply
\begin{equation}
\label{EOM:Q}%%%
i\dot{\mathcal{Q}}_{\gamma}(\bm{\kappa}) = \mathcal{J}_{\gamma}(\bm{\kappa}),
\end{equation}
where the channel flux is $\mathcal{J}_{\gamma}(\bm{\kappa})=D_{\gamma\mu}J_\mu(\bm{\kappa})$.
The surface flux will eventually vanish as $t\rightarrow\infty$, since the outgoing wavepacket passing through the surface eventually vanishes. Therefore, $\dot{\mathcal{Q}}_\gamma$ also approaches zero in this limit, ensuring that $\mathcal{P}_\gamma$ satisfies Eq.~\eqref{Condition:Conservation}. We may point out that $\mathcal{Q}_\gamma(\bm{\kappa})$ are the time-dependent Dyson orbitals in momentum representation, a natural generalization of static Dyson orbitals. Motivated by this observation, we identify that the state associated with a time-dependent Dyson orbital, $|\psi_\gamma\rangle\equiv D_{\gamma\mu}|\phi_\mu\rangle$, satisfies $i|\dot{\psi}_\gamma\rangle=\hat{F}|\psi_\gamma\rangle$. Since MCTDHF coincides with the \textit{exact} ME-TDSE in the limit of an infinite number of orbitals, the formalism sheds light on the existence of \textit{exact} EOMs for the time-dependent Dyson orbitals, a fundamental property of ME-TDSE beyond MCTDHF, and provides a systematic way to approach it.

To illustrate the effectiveness of our formulation and clarify misinterpretations based on instantaneous natural orbitals, we first perform numerical simulations for a prototypical system: hydrogen anion $\mathrm{H^-}$.
As one of the simplest two-electron systems, $\mathrm{H^-}$ exhibits strong ground-state correlation, giving rise to a considerable probability of multichannel ionization~\cite{rau1996negative,andersen2004atomic}. We consider single-photon ionization of $\mathrm{H}^-$ driven by a linearly polarized, $10$-cycle sine-square laser pulse with central frequency $\omega=1$~a.u. The minimal single and double ionization threshold is $0.0277$~a.u.~and $0.5277$~a.u., respectively. By absorption of a $\omega=1$~a.u. photon, the residual H atom may be left in the ground state, excited states, or continuum. For clarity and computational efficiency, we focus on the $1s$ and $2s$ ionization channels.

\begin{figure}[htb!]
\includegraphics[width=0.5\textwidth]{\detokenize{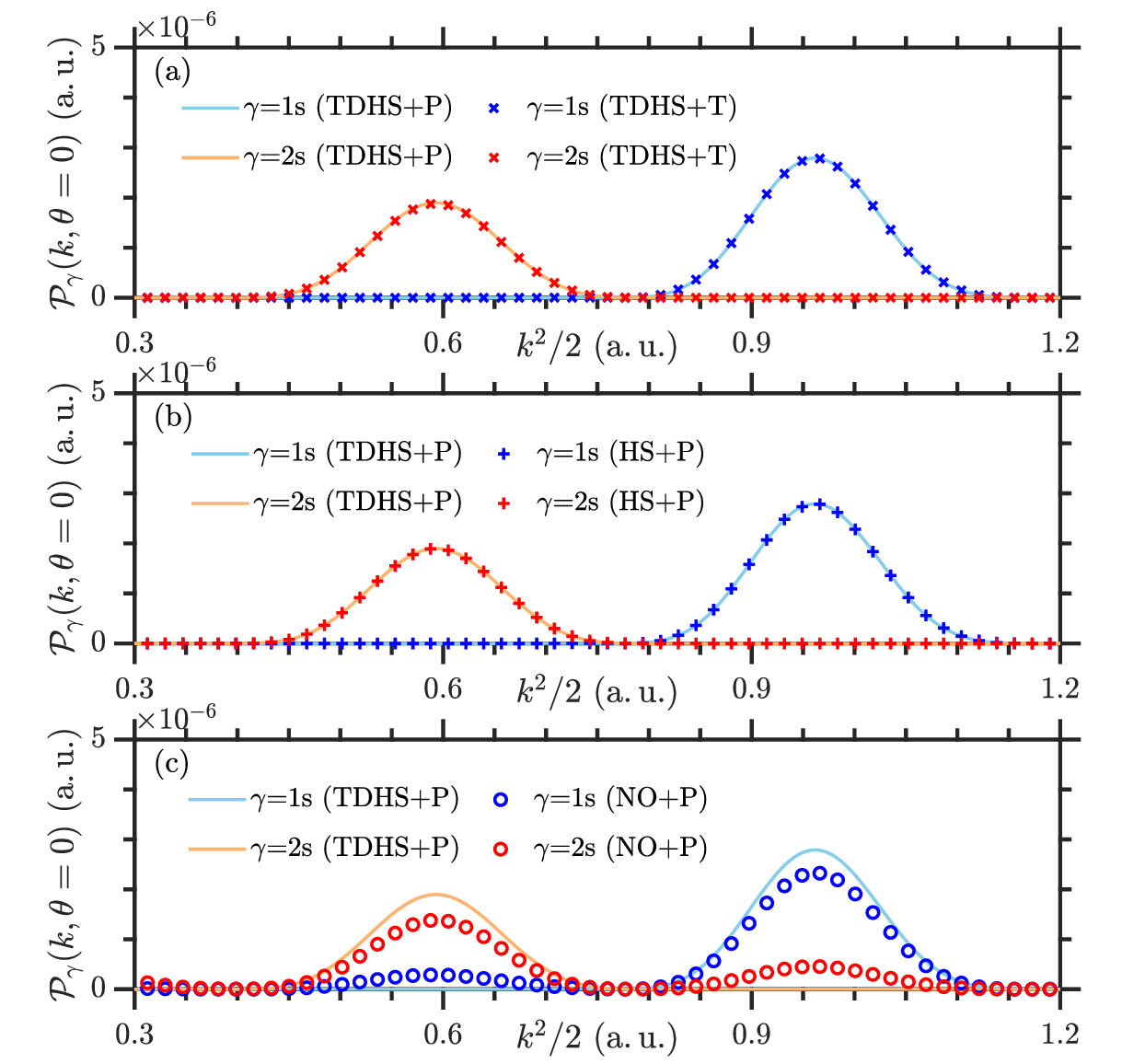}}
\caption{ (colored online) The channel-resolved PMD of $\mathrm{H^-}$ along the laser polarization by different approaches. See texts for laser parameters.
(a) Comparison of TDHS+P and TDHS+T.
(b) Comparison of TDHS+P and HS+P.
(c) Comparison of TDHS+P and NO+P. 
See the legends for interpreting the plots. The results are extracted from a calculation with a two-electron complete active space constituting 31 orbitals.}
\label{FigComparePMD}
\end{figure}

We compare the channel-resolved PMDs at $\theta=0$ (parallel to laser polarization) as functions of photoelectron energy ($k^2/2$), extracted using three different choices of $|\gamma\rangle$:~(\romannumeral 1) TDHS, with $D_{\gamma\mu}(t)$ from Eq.~\eqref{EOM:D};~(\romannumeral 2) field-free HS $|\gamma(0)\rangle$ determined by Eq.~\eqref{Koopmans}, with 
\begin{equation}
D_{\gamma\mu}(t) = (Z^\dag(0))_{\gamma\nu}\langle\Psi(0)|\hat{a}^\dag_\nu(0) \hat{a}_\mu(t)|\Psi(t)\rangle;
\end{equation}
(\romannumeral 3) instantaneous natural orbitals (NO), where the eigen decomposition $\rho_{\mu\nu}(t)=u_{\mu\gamma}(t)\lambda_\gamma(t)u^*_{\nu\gamma}(t)$ yields
\begin{equation}
D_{\gamma\mu}(t)=u^*_{\mu\gamma}(t)\lambda^{1/2}_\gamma(t).
\end{equation}
Figure~\ref{FigComparePMD}(a) shows results obtained with TDHS using projection (P) and time-dependent surface flux (T) approaches, whose details are given in \textit{Appendix A}. In each channel, a single peak appears at the predicted positions ($0.9723$ and $0.5973$ a.u.\ for $1s$ and $2s$, respectively), and the two methods perfectly coincide.
Since hole dynamics are negligible in this case, using the field-free HS as $\gamma$ yields essentially identical results, as seen in Fig.~\ref{FigComparePMD}(b). In contrast, employing instantaneous NOs to construct the $(N-1)$-electron reference states produces two peaks in each channel [Fig.~\ref{FigComparePMD}(c)], with nearly equal contributions from the two channels. This behavior directly signals the inadequacy of NOs in approximating Dyson orbitals, and highlights the potential pitfalls of the NO-based approach in multichannel ionization~\cite{ohmura2020analysis,ohmura2022investigation}.

Having addressed the conceptual challenge of defining channel-resolved observables, we now illustrate how the formalism brings new physical insights. With well-defined TDHSs, we show that the $\omega-2\omega$ laser~\cite{prince2016coherent,di2019complete,you2020new} enables a novel coherent control strategy for hole dynamics in the photoionization process of dimer systems, i.e., $\mathrm{X_2}\longrightarrow\mathrm{X_2^+}$, where $\mathrm{X}$ is a monomer species. Following single ionization, the created hole could occupy a mixture of the monomers' hole states and subsequently hop between the monomers after the laser is switched off. Because the two monomers are well separated, the cation $\mathrm{X_2^+}$ supports two quasi-degenerate eigenstates that are approximately the symmetric and antisymmetric combinations of localized hole configurations, e.g., $|\phi^{-1}_\pm\rangle\propto|\phi^{-1}_L\rangle\pm|\phi^{-1}_R\rangle$. The hole dynamics is fully characterized by the ion reduced density matrix (iRDM), obtained in our formalism by tracing over the photoelectron degrees of freedom,
\begin{equation}
    \rho^{\mathrm{(i)}}(\gamma,\gamma')=\frac{1}{\mathcal{N}} \SumInt_{\bm{\kappa}} \mathcal{Q}_{\gamma}(\bm{\kappa})\mathcal{Q}^*_{\gamma'}(\bm{\kappa}),
    \label{Def:iRDM}
\end{equation}
where $\gamma,\gamma'=\phi^{-1}_\pm$ and $\mathcal{N}$ is a normalization factor to ensure $\mathrm{Tr\rho^{\mathrm{(i)}}}=1$. The iRDM has a transparent physical interpretation that its eigenvectors characterize the charge eigenstates of the ion, and the eigenvalues give their populations. After the laser pulse, non-zero off-diagonal elements of the iRDM indicate that $\mathrm{X_2^+}$ remains in a coherent superposition of $|\phi^{-1}_\pm\rangle$, leading to hole hopping between the two monomers with a period determined by their small energy splitting. The larger the magnitude of the coherence term, the more pronounced the charge migration is.

The bichromatic $\omega-2\omega$ laser is particularly effective in creating and controlling such coherence because of its symmetry-breaking ability via multi-pathway ionization:
\begin{subequations}
\begin{align}
\label{Processes:1}
    \mathrm{X_2}+1\times(2\omega)&\longrightarrow \mathrm{X_2^+}(\phi^{-1}_+)+\mathrm{e^-}(\text{parity}-),\\
\label{Processes:2}
    \mathrm{X_2}+2\times(\omega)&\longrightarrow \mathrm{X_2^+}(\phi^{-1}_-)+\mathrm{e^-}(\text{parity}-),\\
\label{Processes:3}
    \mathrm{X_2}+1\times(2\omega)&\longrightarrow \mathrm{X_2^+}(\phi^{-1}_-)+\mathrm{e^-}(\text{parity}+),\\
\label{Processes:4}
    \mathrm{X_2}+2\times(\omega)&\longrightarrow \mathrm{X_2^+}(\phi^{-1}_+)+\mathrm{e^-}(\text{parity}+).
\end{align}
\label{Processes}
\end{subequations}
Unlike monochromatic fields, the interchannel integral [Eq.~\eqref{Def:iRDM}] for $(\gamma,\gamma')=(\phi^{-1}_+,\phi^{-1}_-)$ is nonzero since the photoelectron in the two ion channels possesses the same parity [e.g., Eqs.~\eqref{Processes:1}\eqref{Processes:2}]. As the energy difference between $|\phi^{-1}_\pm\rangle$ is typically small ($\sim0.01$~a.u.) compared to the spectral width of the pulse, a large $|\rho^{\mathrm{(i)}}(\phi^{-1}_+,\phi^{-1}_-)|$ may arise due to the considerable overlap between the broad peaks in $\mathcal{Q}_{\phi^{-1}_\pm}(\bm{\kappa})$. As a key advantage of $\omega-2\omega$ pulses, the phase delay $\varphi$ between the components directly tunes the strength of the hole dynamics, since $\rho^{\mathrm{(i)}}(\phi^{-1}_+,\phi^{-1}_-)\propto c_+\exp (i\varphi)+c_-\exp(-i\varphi)$, where $c_\pm$ are composed of individual transition amplitudes of the processes in Eq.~\eqref{Processes}, see \textit{Appendix B}.
We emphasize that this phase-dependent modulation differs fundamentally from the well-known asymmetry-parameter control in photoelectron spectroscopy~\cite{yin1992asymmetric,yamazaki2007observation}, which is based exclusively on intrachannel information.
\begin{figure}[!th]
\includegraphics[width=0.5\textwidth]{\detokenize{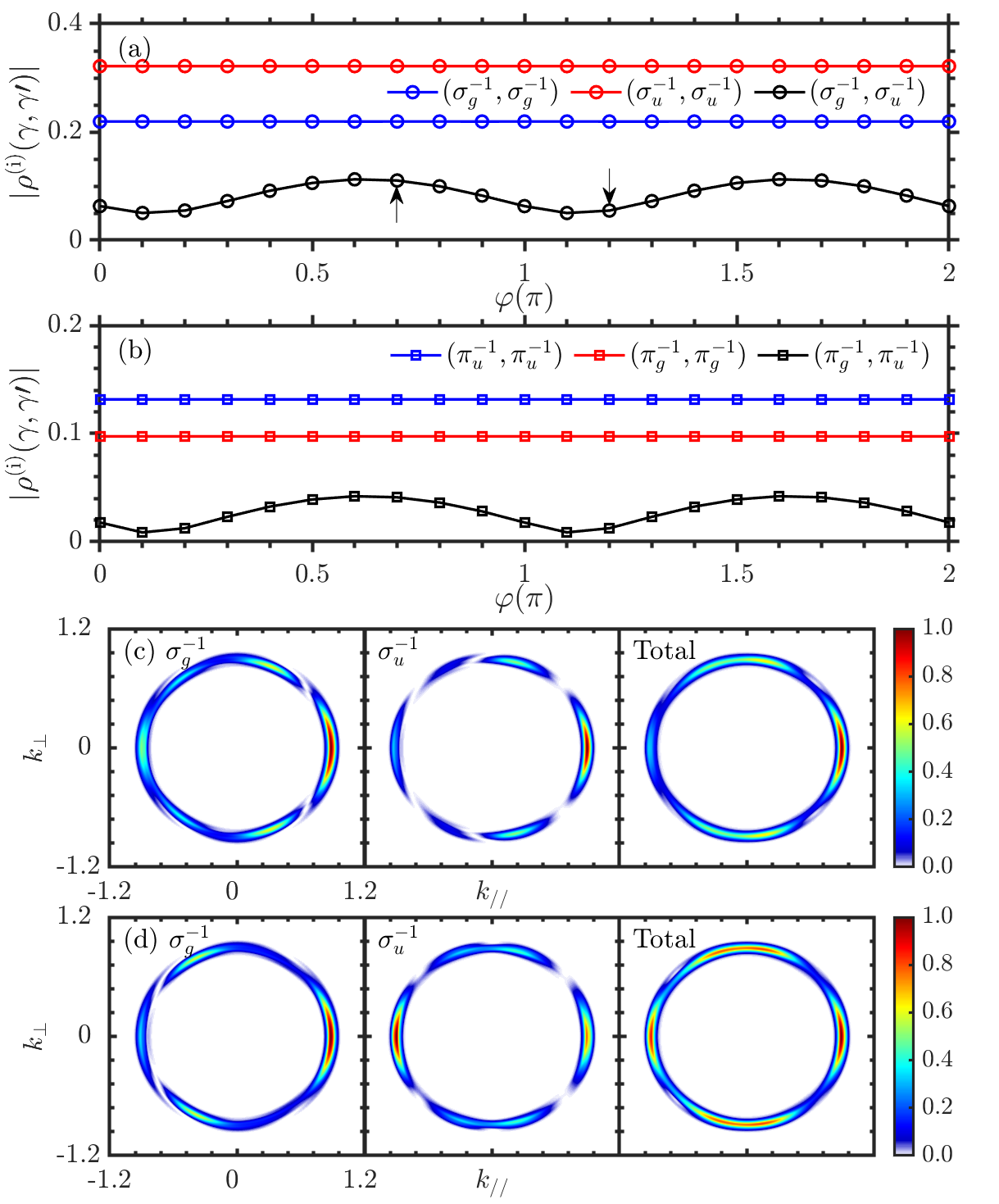}}
\caption{Calculations of $\mathrm{Ne_2}$ with a fixed nuclear distance of 5.841~a.u., driven by $\omega$-$2\omega$ pulses with a fundamental frequency of 0.60~a.u., a sine-square envelope lasting for 5~fs, and a tunable phase delay $\varphi$. The peak intensities for the $\omega$ and $2\omega$ components are $10^{13}$~W/cm$^2$ and $10^{10}$~W/cm$^2$, respectively. (a)-(b) Magnitudes of the iRDM elements as functions of $\varphi$. Entry indices are indicated in the legend. Symmetry forbidden entries are zero and not shown. Channel-resolved and total PMDs at (c)~$\varphi=0.7\pi$ (d)~$\varphi=1.2\pi$. All PMDs are normalized to the individual maximum for clarity, and share the same range of axis. $k_{//}$, $k_\perp$ are parallel and perpendicular components of photoelectron momentum in a.u., respectively. }
\label{FigDimer}
\end{figure}

To validate the proposed $\omega$–$2\omega$ strategy, we perform \textit{ab initio} simulations incorporating the TDHS formalism beyond the limit of sudden-removal~\cite{cederbaum1986correlation,cederbaum1999ultrafast}. For demonstration, we consider $\mathrm{X}=\mathrm{Ne}$ with a fixed internuclear separation of $5.841$~a.u. (see \textit{Appendix C} for justification of the chosen system and \textit{Appendix D} for computational details). We set the $\omega$–$2\omega$ pulses linearly polarized along the molecular axis. Consequently, the single ionization from the valence shell could lead to three independent pairs of dimer hole states,
\begin{equation}
    \begin{aligned}
        |\sigma^{-1}_{u/g}\rangle &= |(2p_m)^{-1}_{L}\rangle\pm|(2p_m)^{-1}_{R}\rangle \text{ where } m=0, \\
        |\pi^{-1}_{u/g}\rangle &=  |(2p_m)^{-1}_{L}\rangle\pm|(2p_m)^{-1}_{R}\rangle \text{ where } m=\pm1.
    \end{aligned}
\end{equation}
Restricted by symmetry, the full $6\times6$ iRDM decomposes into three $2\times2$ blocks, each governing the hole dynamics in the $\sigma$- and $\pi$-mode. Figure~\ref{FigDimer}(a) and~\ref{FigDimer}(b) present the magnitudes of the diagonal and non-zero off-diagonal elements, $|\rho^{\mathrm{(i)}}(\gamma,\gamma')|$, as functions of the relative phase $\varphi$ between pulses. As expected, the diagonal entries are insensitive to $\varphi$, while the off-diagonal entries exhibit pronounced $\pi$-periodic oscillations, directly demonstrating phase-controlled modulation of charge migration. To gain deeper intuition, we recall Eq.~\eqref{Def:iRDM} and analyze the interchannel photoelectron overlap between the $\sigma^{-1}_g$ and $\sigma_u^{-1}$ channels at two representative phases, $\varphi = 0.7\pi$ and $1.2\pi$, corresponding to maximally and minimally coherent, respectively, as shown in Figs.~\ref{FigDimer}(c) and~\ref{FigDimer}(d). At $\varphi=0.7\pi$, the photoelectron is predominantly emitted towards the right side in both $\sigma$ channels, resulting in an enhancement of interchannel overlap. In contrast, at $\varphi=1.2\pi$, the overlap is reduced due to the opposite direction of emission in $\sigma^{-1}_u$ and $\sigma^{-1}_g$. This comparison demonstrates a key advantage of the TDHS framework. Without channel resolution, only the total PMD would be observable, concealing the essential interchannel structure that encodes the underlying hole dynamics.

In conclusion, we have developed a time-domain generalization of the extended Koopmans’ theorem, fully consistent with the MCTDHF framework. This advances the long-standing challenge of disentangling correlated electron-hole dynamics by providing a rigorous and transparent definition of time-dependent hole states. 
According to TDHS theory, a two-color strategy is carried out to steer the hole in a dimer system after its single ionization. 
The power of the formulation is illustrated through different mechanisms for controlling the hole dynamics, 
demonstrating its potential as a versatile tool for exploring ultrafast multielectron dynamics.

\section*{acknowledgment}
This work was supported by the National Natural Science Foundation of China (NSFC) (Grant Nos. 12450405, 12274294), the Fundamental Research Funds for the Central Universities (Grant No. AF0720103).
This research was also supported in part by a Grant-in-Aid for Scientific Research (Grant No. JP19H00869, JP20H05670, JP22H05025, JP24H00427, and JP25K01688) from the Ministry of Education, Culture, Sports, Science and Technology (MEXT) of Japan, and the RIKEN TRIP initiative. This research was also partially supported by the MEXT
Quantum Leap Flagship Program, (Grant No. JPMXS0118067246)
and JST COI-NEXT, (Grant No. JPMJPF2221).
The computations in this paper were run on the Siyuan-1 cluster supported by the Center for High Performance Computing at Shanghai Jiao Tong University.
%=================================================================================================================
\section*{Data availability}
Data supporting the findings of this work are openly available \cite{data}.

\bibliography{reference}

\section*{End Matter}\label{EndMatter}
\textit{Appendix A: Two approaches for channel-resolved momentum distributions in MCTDHF}--Beyond its conceptual significance, our generalization also introduces practical numerical utilities. With a readily solvable EOM of $D_{\gamma\mu}$ [Eq.~\eqref{EOM:D}], Eq.~\eqref{EOM:Q} naturally yields a channel-resolved time-dependent surface flux (t-SURFF) approach in MCTDHF. The central ingredient, the surface flux $J_\mu(\bm{\kappa})$, has already appeared in the standard single-electron t-SURFF approach~\cite{zhang2023qpc} and in conventional MCTDHF t-SURFF implementations~\cite{orimo2019application}. The only extra effort is to solve $D_{\gamma\mu}$. Furthermore, Eq.~\eqref{Def:Qgamma} provides a direct channel-resolved projection approach, since $\phi_i(\bm{\kappa})$ is readily obtained by projecting orbitals onto continuum states such as masked plane waves or Coulomb waves.

In the example of $\mathrm{H}^-$, the convergence of channel-resolved t-SURFF and projection approaches is tested by enlarging the mask radius. We have numerically verified the invariance of $\mathcal{P}_\gamma$ against the choice of gauge.\\

\textit{Appendix B: A derivation of the modulation pattern of off-diagonal elements of iRDM}--The oscillation behavior in $|\rho^{\mathrm{(i)}}(\phi^{-1}_+,\phi^{-1}_-)|$ can be understood as follows
\begin{equation}
    \begin{aligned}
    \SumInt_{\bm{\kappa}}\mathcal{Q}_{\phi^{-1}_+}(\bm{\kappa})\mathcal{Q}^*_{\phi^{-1}_-}(\bm{\kappa}) &= c_+ e^{i\varphi} + c_- e^{-i\varphi},    
    \end{aligned}
\end{equation}
where the $\varphi$-independent coefficients are
\begin{subequations}
\begin{align}c_{-}&=\SumInt_{\bm{\kappa}}\mathcal{Q}^{(2\omega)}_{\phi^{-1}_-}(\bm{\kappa})\mathcal{Q}^{(1\omega)*}_{\phi^{-1}_+}(\bm{\kappa}),\\
c_{+}&=\SumInt_{\bm{\kappa}}\mathcal{Q}^{(1\omega)}_{\phi^{-1}_-}(\bm{\kappa})\mathcal{Q}^{(2\omega)*}_{\phi^{-1}_+}(\bm{\kappa}).
\end{align}
\label{def:cplusminus}
\end{subequations}
Here, $\mathcal{Q}^{(1\omega)}_\gamma(\bm{\kappa})$ corresponds to the transition induced by two-photon absorption of $\omega$ photons [Eqs.~\eqref{Processes:2}\eqref{Processes:4}], while $\mathcal{Q}^{(2\omega)}_\gamma(\bm{\kappa})$ corresponds to single-photon absorption of $2\omega$ photons [Eqs.~\eqref{Processes:1}\eqref{Processes:3}] when $\varphi=0$. They are the process-specific transition amplitudes discussed in the main text. The oscillation pattern differs from that of photoelectron asymmetry since the interchannel integrals like Eq.~\eqref{def:cplusminus} never appear in the latter.\\

\textit{Appendix C: On the reality of neon dimer example}--For experimental feasibility, the choice of dimer species $\mathrm{X}_2$ should minimize nuclear motion effects that can wash out the ultrafast hole dynamics. Heavy noble-gas dimers ($\mathrm{Ne_2}$, $\mathrm{Ar_2}$, $\mathrm{Kr_2}$) and alkali-metal dimers ($\mathrm{Na_2}$, $\mathrm{K_2}$, $\mathrm{Rb_2}$) are natural candidates, as their large reduced masses lead to slow nuclear dynamics and localized nuclear wavepackets. As a practical balance between computational complexity and experimental relevance, we select $\mathrm{Ne_2}$ for demonstration. Spectroscopy studies~\cite{wuest2003determination} report an equilibrium internuclear distance of $R_\mathrm{Ne-Ne}=(3.094\pm0.001)$~\AA, and the lowest vibrational state has a spatially localized wavepacket that spans approximately $2.5\sim4.0$~\AA. The comparatively weak nuclear averaging is consistent with the clear double-slit interference observed in strong-field ionization of $\mathrm{Ne_2}$~\cite{kunitski2019double}. Crucially, the hole-migration period inferred from the energy splitting between $\mathrm{Ne^+_2}(^2\Sigma_u^+)$ and $\mathrm{Ne^+_2}(^2\Sigma_g^+)$ at $R_\mathrm{Ne-Ne}$~\cite{ha2003lowest} is $\sim20$~fs. Furthermore, the nuclear motion in $\mathrm{Ne_2}$ evolves over timescales exceeding $\sim100$~fs due to its large mass. Therefore, the hole dynamics can persist long enough to be observed before appreciable decoherence from nuclear motion occurs. These considerations should justify the fixed-nuclei approximation in the present work.\\

\textit{Appendix D: Computational details of $\omega$-$2\omega$ example}--Both the $\omega$ and $2\omega$ pulses are linearly polarized along the molecular axis, with vector potentials 
\begin{subequations}
\begin{align}
    A^{(1\omega)}(t) &= A_{\max}^{(1\omega)}f(t)\sin(\omega t),\\
    A^{(2\omega)}(t) &= A_{\max}^{(2\omega)}f(t)\sin(2\omega t+\varphi),
\end{align}    
\end{subequations}
where $\omega=0.6$ a.u. is the fundamental frequency. The sine-squared envelope $f(t)$ has a total foot-to-foot duration of $2M\pi/\omega$ where $M=20$ (about 5~fs). The intensities are fixed at $\mathrm{10^{13}\,W/cm^2}$ for the $\omega$-pulse and $\mathrm{10^{10}\,W/cm^2}$ for the $2\omega$-pulse, respectively. The experimentally tunable parameter is $\varphi$, which sets the relative phase delay between the two pulses and breaks the reflection symmetry.

We fix $R_{\mathrm{Ne}-\mathrm{Ne}}$ at $5.841$~a.u. Among all 20 electrons, the lowest 8 electrons occupying 1s and 2s are frozen, and the remaining 12 electrons are fully active in three dimensions. We employ the spin-restricted formalism and a complete active space spanned by 14 orbitals, which has $\sim1,700,000$ symmetry-allowed configurations (determinants) in total. The time-dependent orbitals $\phi_\mu(\boldsymbol{r},t)$ are treated with single-center expansions based on spherical harmonics and radial B-spline functions, where the maximal angular momentum is 100, and the simulation region extends up to $r=96$~a.u. An absorptive potential removes wavepackets surpassing $r=64$~a.u. An \textit{ab initio} effective core potential is adopted to replace the frozen 1s electrons on both Ne atoms~\cite{nicklass1995ab}, which significantly improves numerical efficiency with minor compromise on the accuracy.

The MCTDHF equations are solved by a fourth-order exponential time-differencing Runge-Kutta solver with a fixed time step of $0.002$~a.u.
The iRDM elements are verified to converge after sufficient post-pulse propagation. Due to cylindrical symmetry, the PMDs are presented in two dimensions. Details of our implementation are found in the 
accompanying work~\cite{zhang2025time}.

In our real-time simulation, the ionization energy of the four highest orbitals ($\sigma_u$, $\sigma_g$, $\pi_u$, $\pi_g$) are $0.8027$, $0.8106$, $0.8007$ and $0.8018$~a.u., respectively, reporting an energy difference of 0.213~eV for $\sigma$ and 0.031~eV for $\pi$. We also calculate the same energy differences with a more precise quantum chemistry approach, IP-EOM-CCSD, using the implementation in PySCF and the aug-cc-pV6Z basis. The reported values are 0.234~eV and 0.032~eV, which is consistent with our results, indicating that our calculation remains predictive in a realistic scenario. We mention that, due to the complete neglect of correlation effects, the absolute values of the ionization energies in a Hartree-Fock calculation are systematically higher than in IP-EOM-CCSD by 0.06~a.u., while in our multiconfigurational treatment with the inclusion of $3s3p$ orbitals, this error is reduced to 0.015~a.u.\\

%\begin{thebibliography}{100}
%\end{thebibliography}

\end{document}